\documentclass[aps,prb,twocolumn,groupedaddress,eqsecnum]{revtex4}

\usepackage{amsfonts}
\usepackage{dcolumn}
\usepackage{graphicx,amssymb,amsmath}
\usepackage{bm}
\usepackage{natbib}
\usepackage{epstopdf}
\begin{document}

\title{ Statistical, collective and critical phenomena of classical one-dimensional disordered Wigner lattices}

\author{Shimul Akhanjee}
\email[]{shimul@physics.ucla.edu}
\author{Joseph Rudnick}
\email[]{jrudnick@physics.ucla.edu}
\affiliation{Department of Physics, UCLA, Box 951547, Los Angeles, CA 90095-1547}


\date{\today}

\begin{abstract}
We explore various properties of classical one-dimensional Wigner solids in the presence of disorder at $T=0$ in the context of a recently discovered Anderson transition of plasma modes in the random potential system. The extent to which the Wigner lattice is really a ``crystal'' rather than an amorphous solid is discussed for two types of disorder. We investigate the way in which zero point quantum fluctuations that would normally destroy the long range positional order are affected by the disorder. The probability density of particle spacings is examined analytically within a weak disorder approximation and compared to numerical calculations for two different realizations of disorder. We also discuss the multifractal character of the eigenmodes, the compressibility of the electrons, and the AC conductivity.

\end{abstract}


\maketitle

\section{Introduction}

In 1958, E. Wigner predicted that an electron gas in a positive jellium background would, at low density, solidify into an ordered lattice\cite{wigner:1d_wc}. Moreover, in classical systems that have been studied in certain soft condensed matter models, the strong Coulomb repulsion is sufficient to cause a system of particles to crystallize \cite{boris}.

In such a phase, the most natural physical quantities to consider are the elementary excitations. The charged collective modes or plasmons at $T=0$ are entities of this sort, the essential features of which are captured by a classical description\cite{giampwc:1d_wc}. It is important to emphasize that the inclusion of strong disorder either chemically or, in the case of a low dimensional system, via a rough external substrate will result in the destruction of a well-defined base lattice. That is, the term ``Wigner crystal'' is no longer entirely descriptive in the presence of sufficiently strong disorder. Rather, that system is closer to a ``glass'' than a crystalline solid. 

Formally, the existence of Bragg peaks indicates that the system is a solid. However, others have discussed the possibility that the system retains quasi-long range order as a consequence of disorder\cite{bragg}. Subtle implications are entailed by the term ``glass'' in describing a system in terms of its magnetic and topological properties. Here, we restrict our definition of ``glass-like''  to a system with a random positional configuration at zero magnetic field, even though a 1D system cannot truly be topologically disordered as in higher dimensional systems \cite{zimandisorder}.  
	
In this article we focus on two aspects of one dimensional disordered Wigner crystal(1DWC) systems---in particular the plasma oscillations and the statistical behavior of the equilibrium electron locations. We numerically study the plasma modes of the disordered system and calculate the compressibility and the AC conductivity. Regarding the structural properties, we consider whether the distribution of the spacings between adjacent electrons can be analytically understood in terms of the randomness in other parameters in the Hamiltonian. A mathematical formalism is developed that provides some insight into the connection between power spectrum of the randomness and the positional statistics.
We also attempt to address the long-standing issue of whether a classical Wigner crystal in the presence of either random field or random charge disorder preserves some degree of crystalline integrity. 

\section{The Statistical Configuration of the Particles Under the Influence of Weak Disorder}

Let us now consider how disorder can affect the positional ordering of the particles in the 1DWC system. Evidently, the nature of disorder in such a system is not unique.  It can exist in the spacings between the particle sites, the masses, the effective interaction between particles, or one can have a set of charges pinned by a random potential. It is also the case that the introduction of randomness in one parameter can induce randomness in another. For example, randomness in particle-particle interactions or an external field may well result in the destruction of a periodic base lattice at mechanical equilibrium. 

The ordered system of size $L$ is defined by the Hamiltonian,
\begin{equation}
H = \sum\limits_{i = 1}^L {\frac{{p_i^2 }}{{2m_e^* }} + \frac{1}{2}} \sum\limits_{i \ne j}^{} {\frac{{e_0^2 }}{{\left| {x_i  - x_j } \right|}}}
\label{eq:plashamilt} 
\end{equation}
where the quantities $p_i$ and $m_e^*$ refer to the momentum and effective mass of the  $i^{\rm th}$ particle, respectively. Our system lives in the low density regime in which the Coloumb interactions are much more influential than the kinetic energy, resulting in the crystalline ordering of the particles with the spatial coordinates $\lbrace x_i \rbrace$ separated with lattice constant $a$. Moreover, the system is stabilized by a positive Jellium neutralizing background. We wish to examine the vibrational modes of the system taking into account the full electrostatic force acting on each charge.

One can add to this Hamiltonian a term representing the interaction of the charges with a random electrostatic potential $V(x)$, defined as,
\begin{equation}
V(x) = J\sum\limits_n^N {a_n } \cos (2\pi nx) + b_n \sin (2\pi nx)
\label{eq:rpfourier}
\end{equation}
where the parameter $J$ is a dimensionless coupling constant. The random variables $a_n$ and $b_n$ have a mean equal to zero. Moreover, their distributions are controlled by independent Gaussian distributions the widths of which are independendent of $n$, yielding a "white noise" power spectrum. We will call this system Model A.

Additionally, we can explicitly inject randomness into the single charge values $Q_i$. We term this random charge system Model B, where the Hamiltonian is obtained from (\ref{eq:plashamilt}) by replacing the $e_0^2$ with $Q_i \times Q_j$.  Some possible realizations of this model may be found in soft condensed matter systems such as in biopolymer arrays that have contain randomly assigned chemical constituents and unscreened Coulomb interactions\cite{boris}. 

In both Model A and Model B, the charges in our lattice will relax to mechanical equilibrium, at locations that differ from the evenly-spaced positions they occupy in the ordered case.  Here, we characterize key properties of the equilibrium configuration, focusing on the statistical characteristics of the spacings between neighboring points. 
Finally, we discuss the extent to which the resulting lattice exhibits crystalline order.

As a first step in answering these questions for Model B, we examine the product density functions for different distributions of charges. This is shown in Appendix \ref{sec:appA}. There is a direct relationship between this distribution and the distribution of forces in the case of strongly-screened interactions. The general distribution of the forces on a particular  charge when interactions are unscreened requires a more involved analysis. Given this probability distribution,  we turn to the distribution function of the spacings between the charges at equilibrium. It turns out that an analytical result that is valid in the weak disorder regime---based on the assumption that the width of the single charge distributions is very small---agrees remarkably well with the results of numerical calculations. We will call this result the ``viscosity approximation'' in that it is based on the notion that the charges  will deviate from the ordered lattice by distances that are proportional to the total force acting on each charge $F_i^{tot}  \propto  \Delta x_i$ when they are equally spaced. That is,  $P(F_i^{tot})  \propto   P(\Delta x_i)$, as if the system were placed in a highly viscous medium and initial displacements of the charges from their initial positions were predictive of their eventual positions. Given this assumption, one can work out the solution to Model B for all three of the distributions noted above starting with the gaussian system. 

The distribution of forces on the  $i^{\rm th}$ charge is given by the convolution:
\begin{equation}
\begin{aligned}
P_F (F_i ) &= \int\limits_{ - \infty }^\infty  {P_E (E_i )} P_q (q_i )\delta (Eq - F_i )dE_i dq_i\\
           &= \int\limits_{ - \infty }^\infty  {P_E ({\raise0.7ex\hbox{${F_i }$} \!\mathord{\left/{\vphantom {{F_i } {q_i }}}\right.\kern-\nulldelimiterspace}
\!\lower0.7ex\hbox{${q_i }$}})} P_q (q_i )\frac{{dq_i}}{{| q_i |}} 
\end{aligned}
\label{eq:convforce}
\end{equation} 
We must first determine $P_E (E_i)$, of which describes the likelihood that a given charge will have an electric field acting on it. The electric field acting on the  $i^{\rm th}$  charge is $F^i _{total}  = \sum\limits_{i \ne j} {q_j \xi _{i,j} }$, where $\xi _{ij}  \approx \frac{1}{{(x_i  - x_j)^2 }}$.  Subsequently, if we examine the first charge in the array, the distribution of electric fields acting on it is a distribution governed by the convolution:
\begin{equation} 
\begin{aligned}
P_E (E_1 ) &= \int\limits_{ - \infty }^\infty  {dq_2 } \int\limits_{ - \infty }^\infty  {dq_3 } \cdots \int\limits_{ - \infty }^\infty  {dq_N } P_q (q_2 )P_q (q_3 ) \cdots P_q (q_N )\\
&\times  \delta (q_2 \xi _{2,1}  + q_3 \xi _{3,1}  +  \cdots  + q_N \xi _{N,1}  - E_1 ) 
\end{aligned}
\label{eq:econv}
\end{equation}
This convolution can be simplified by making use of the Fourier representation of the Dirac delta function,
\begin{eqnarray}
\lefteqn{\delta (q_2 \xi _{2,1}  + q_3 \xi _{3,1}  +  \cdots  + q_N \xi _{N,1}  - E_1 )} \nonumber \\ 
&= \frac{1}{{2\pi }}\int\limits_{ - \infty }^\infty  {\exp [i\Omega 
(} q_2 \xi _{2,1}  + q_3 \xi _{3,1}  +  \cdots  + q_N \xi _{N,1}  - E_1 )]d\Omega \nonumber \\ 
\label{eq:fourdelta}
\end{eqnarray}
More generally, for any given charge, the distribution of the electric fields $P_E (E_i)$ is transformed into the following single integration,
\begin{equation}
P_E (E_i ) = \frac{1}{{2\pi }}\int\limits_{ - \infty }^\infty  {e^{ - i\Omega E_i } } \left( {\prod\limits_{k \ne i}^N {g{}_k(\Omega )} } \right)d\Omega 
\end{equation}
where, we have integrated out the charges by performing an integral that defines $g(k)$,
\begin{equation}
\begin{aligned}
&g{}_k(\Omega ) = \int\limits_{ - \infty }^\infty  {P_q (q_k } )\exp [i\Omega q_k \xi _{k,i} ]dq_k \\ 
&= \exp [i\Omega \xi _{k,i} q_0  - 
\xi _{k,i}^2 \Omega ^2 \sigma /2]
\end{aligned}
\label{eq:gfunc}
\end{equation}

The resulting distribution of electric fields is in closed form,

\begin{widetext}
\begin{equation}
P_E (E_i ) = \frac{1}{{2\pi }}\int_{ - \infty }^\infty  {\exp \left[ {i\Omega \left( {\sum\nolimits_{k = 1}^N {\xi _{k,i} q_0  - } E_i } \right) - \frac{{\Omega ^2 \sigma }}{2}\sum\nolimits_{k \ne i}^N {\xi _{k,i}^2 } } \right]} d\Omega
= \frac{{\exp \left( { - \left( {\sum\nolimits_{k = 1}^N {\xi _{k,i} q_0  - } E_i } \right)^2 /\left( {2\sigma \sum\nolimits_{k \ne i}^N {\xi _{k,i}^2 } } \right)} \right)}}{{\left( {2\pi \sigma \sum\nolimits_{k \ne i}^N {\xi _{k,i}^2 } } \right)^{1/2} }}
\label{eq:gdistb}
\end{equation}
\end{widetext}

The distribution of the force can now be determined by using (\ref{eq:convforce}). As a consequence of the periodic boundary conditions we can take,
$\sum_{k \ne i}^N \xi _{k,i}  = 0$
Moreover, we can apply the exact sum $
\sum_{r = 1}^{\infty}  1/r^2  = \pi^2/6
$ in the limit of a large system size to the $\xi_{k,i}^2$ summation, reducing (\ref{eq:gdistb}) to a Gaussian distribution centered at zero:
\begin{equation}
P_E (E_i ) = \frac{{\exp \left( { - 3\left( {E_i } \right)^2 /\left( {\pi^2 \sigma } \right)} \right)}}{{\left( {\pi ^3 \sigma /3} \right)^{1/2} }}
\end{equation}
The resulting  force distribution is simply a convolution of two Gaussian  distributions,
\begin{equation}
P_F (F_i ) = \int\limits_{ - \infty }^\infty  {\frac{{\exp \left( { - 3\left( {F_i/q_i } \right)^2 /\left( {\pi \sigma } \right) - 
(q_i  - q_0 )^2 /2\sigma } \right)}}{{\left( {\pi ^2 \sigma /3} \right)^{1/2} \left| {q_i } \right|}}dq_i }
\label{eq:gfinalb}
\end{equation}

We now turn to the other distribution functions. The system with uniform random charges, defined by the density function (\ref{eq:unidist}) can also be evaluated. We can apply convolution of Eq.  (\ref{eq:convforce}). Similar steps lead to the analysis of Eq.  (\ref{eq:gfunc}),
\begin{equation}
\begin{aligned}
& g_k(\Omega)  = \int\limits_{W_1 }^{W_2 } {\frac{{\exp [i\Omega q_k \xi _{k,i} ]}}{{W_2  - W_1 }}}dq_k \\
& = \frac{{2\exp [i\Omega \xi _{k,i} (W_2  + W_1 )/2]\times \sin [\Omega \xi _{k,i} (W_2  - W_1 )/2]}}{{(W_2  - W_1 )\Omega \xi _{k,i} }}\\
\end{aligned}
\end{equation}
After applying equation(\ref{eq:econv}) we have:

\begin{equation}
\begin{aligned}
&P_E (E_i ) = \frac{1}{{2\pi }}\int\limits_{ - \infty }^\infty  {e^{ - i\Omega E_i } }\times\\
&\left( {\prod\limits_{k \ne i}^N {\frac{{e^{i\Omega \xi _{k,i}
(W_1  + W_2 )/2} \sin [\Omega \xi _{k,i} (W_2  - W_1 )/2]}}{{\Omega \xi _{k,i} (W_2  - W_1 )/2}}} } \right)d\Omega 
\end{aligned}
\end{equation}

Similarily, for binary disorder we have
\begin{equation}
\begin{aligned}
&P_E (E_i ) = \frac{1}{{2\pi }}\int\limits_{ - \infty }^\infty  {e^{ - i\Omega E_i } }\times\\ 
&\left( {\prod\limits_{k \ne i}^N {e^{i\Omega \xi _{k,i}
(Q_1  + Q_2 )/2} \cos [\Omega \xi _{k,i} (Q_2  - Q_1 )/2]} } \right)d\Omega 
\end{aligned}
\end{equation}

This leads to the final force distributions for the uniform and the binary systems respectively as:

\begin{equation}
\begin{aligned}
&P_F^{\rm uni} (F_i ) = \int\limits_{ - \infty }^\infty  {\frac{{Ei\left( { - i\Omega F_i /W_1 } \right) - Ei\left( { - i\Omega F_i /W_2 } \right)}}{{2\pi (W_2  - W_1 )}}} \\
&\times \left( {\prod\limits_{k \ne i}^N {\frac{{e^{i\Omega \xi _{k,i}(W_1  + W_2 )/2} \sin [\Omega \xi _{k,i} (W_2  - W_1 )/2]}}{{\Omega \xi _{k,i} (W_2  - W_1 )/2}}} } \right)d\Omega 
\label{eq:unifinal}
\end{aligned}
\end{equation}

and

\begin{equation}
\begin{aligned}
&P_F^{\rm bin} (F_i ) = \int\limits_{ - \infty }^\infty  {\left( {\frac{{\exp \left( { - i\Omega F_i /Q_1 } \right)}}{{4\pi Q_1 }} + \frac{{\exp \left( { - i\Omega F_i /Q_2 } \right)}}{{4\pi Q_2 }}} \right)}\\
&\times\left( {\prod\limits_{k \ne i}^N {e^{i\Omega \xi _{k,i}(Q_1  + Q_2 )/2} \cos [\Omega \xi _{k,i} (Q_2  - Q_1 )/2]} } \right)d\Omega 
\label{eq:binfinal}
\end{aligned}
\end{equation}

where $Ei(x)$ is the exponential integral defined as:
\begin{equation}
Ei(x) =  - \int\limits_{ - x}^\infty  {\frac{{e^{ - t} dt}}{t}} 
\end{equation}
The density of particle spacings is simply proportional to these force distributions.

This derivation in terms of probability convolutions can similarly be extended to the random potential system of Model A. We previously defined the random potential in Eq.  (\ref{eq:rpfourier}). To simplify the analysis we will only consider the cosine contributions and will perform the analysis for a Gaussian distribution of the random amplitudes. The resulting Fourier decomposition yields the expression,
\begin{equation}
V(x) = \frac{1}{{\sqrt L }}\sum\limits_{n = 1}^L {a_n \cos(} n\pi x/L)
\label{eq:rpshort}
\end{equation}
The force on the $i^{\rm th}$ charge is then given by,
\begin{equation}
F(x_i ) = \left. { - e_0 V'(x)} \right|_{x_i }  = \frac{{e_0 }}{{L^{3/2} }}\sum\limits_{n = 1}^L {a_n (n\pi ) \sin(} n\pi x_i /L)
\label{eq:shortforce}
\end{equation}

It is evident that the source of the randomness is not a coupling constant between particles as was the case in Model B. Rather, we relate probability density of random Fourier amplitudes, $a_n$ ("white noise"), to the particle probability density of successive spacings. The probability convolution for the force is given as:
\begin{widetext}
\begin{equation}
\begin{aligned}
&P(F_i ) = e_0 \int\limits_{ - \infty }^\infty  {da_1 } \int\limits_{ - \infty }^\infty  {da_2 }  \ldots \int\limits_{ - \infty }^\infty  {da_L }P_a (a_1 )P_a (a_2 )\\
&\cdots P_a (a_L )\delta \left( {\frac{\pi }{{L^{3/2} }}\left( {a_1 \sin (\frac{{\pi x_i }}{L})+2a_2 \sin (\frac{{2\pi x_i }}{L}) \ldots La_L \sin (\pi x_i )}-F_i \right)} \right)
\label{eq:conva}
\end{aligned}
\end{equation}
\end{widetext}
As a consequence of Eq.  (\ref{eq:conva}) having the same mathematical structure as (\ref{eq:econv}), one can carry out the same steps that lead to Eq.  (\ref{eq:gdistb}), yielding the expression:
\begin{equation}
P(F_i ) = \frac{{e_0 \exp \left( { - F_i^2 /\left( {2\sigma \sum\nolimits_{n = 1}^L {\left( {\frac{{(n\pi )}}{{L^{3/2} }}\sin (\frac{{n\pi x_i }}{L})} \right)^2 } } \right)} \right)}}{{\sqrt {\left( {2\sigma \sum\nolimits_{n = 1}^L {\left( {\frac{{(n\pi )}}{{L^{3/2} }}\sin (n\pi x_i /L)} \right)^2 } } \right)} }}
\label{eq:rpdistfinal}
\end{equation}

Our final result for the distribution of nearest neighbor spacings of Model A is simply a Gaussian distribution with a width that depends on the relative spatial location, $x_i$ of any given particle. This corresponds to the same Gaussian distribution type as the random amplitudes $a_n$ used to construct the random potential. However, the width differs, as the spacings distribution contains a width that is a function of the precise characteristics of the random potential.

In Model B, one can naively compare all of the derived probability densities of nearest neighbor spacings to the probability densities of force couplings given earlier by Eqs. (\ref{eq:gfinalb}), (\ref{eq:unifinal}), (\ref{eq:binfinal}). The binary system contains three characteristic peaks in both quantities, showing similar behavior. Furthermore, in the Gaussian case both the force couplings and the nearest neighbor spacings follow a convolution of two Gaussian distributions. On the other hand, for the uniform distribution the probability density of nearest neighbor spacings does not follow the logarithmic behavior of the force couplings. Ultimately, this indicates the importance of the distribution type in detailing the equilibrium structural properties of 1DWC systems.

As remarked earlier, the viscosity approximation assumes that the particles will deviate from the ordered lattice by a displacement that is proportional to the total force on each particle when the particles are equally spaced. This approximation seems reasonable when the disorder is relatively weak compared to the interaction strength of the Coloumb force. However, it is necessary test this hypothesis by performing a numerical relaxation of the particles for both Models A and B in finite sized systems. We detail the procedure and compare the numerical results to our previously derived expressions in the next section.

\section{The Numerical Relaxation of the Particles to Equilibrium}

\begin{figure}
\centerline{\includegraphics[height=2.0in]{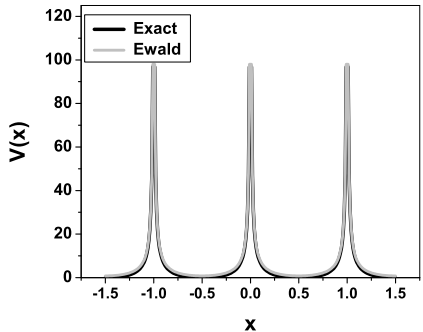}}
\caption{A comparison of the exact and Ewald representations of the Coulomb interaction energy}
\label{fig:ewald}
\end{figure}

The numerical relaxation of a disordered 1DWC to its minimum energy configuration requires the development of two major components. First one must address the complications that arise from performing computations on finite size systems with long ranged interactions. Evidently, the surface effects,  as a result of long range interactions in finite sized systems are not negligible. One possible route to dealing with this complication is to enforce periodic boundary conditions. The standard approach involves the Ewald summation technique\cite{ewald:1d_wc}. We have derived our version of the Ewald potential displayed as Eq.  (\ref{eq:ewaldf}) in Appendix \ref{sec:appB}, that represents the interaction energy between charges on a periodic image, of which is infinitely repeated to assert the periodic boundary conditions. However, the resulting rapidly converging sums still demand computationally intensive efforts. We have discovered that the summation can be reproduced in closed form in terms of a simple function that reproduces the  correct short distance behavior and has the correct periodicity. Consider the lattice summation of the $1/R$ Couloumb potential representing the  infinite sum of n images of length d. The interaction potential between two particles separated by a distance $x$ within a image is given by the expression
\begin{equation}
\phi (x) = \left| {\sum\limits_{n =  - \infty }^\infty  {\frac{{( - 1)^n }}{{(n + x/d)}}} } \right| = \pi \left| {\csc (\pi x/d)} \right|
\label{eq:exactint}
\end{equation}
The result is a simple trigonometric function, of which is easy to implement numerically. Fig.\ref{fig:ewald} shows a graphical comparison of the expressions (\ref{eq:ewaldf}) and (\ref{eq:exactint}).

Next we consider the other required numerical tool. An energy minimization algorithm is neccessary to find the spatial coordinates at which the particles arrange themselves given a particular realization of disorder in the Hamiltonian. These coordinates satisfy a system of linear equations, namely the simultaneous zeros of the total force on each charge. A very powerful technique is the Newton-Raphson(NR) method\cite{press}. In the NR method one recursively computes the Hessian or second derivative matrix and multiplies its inverse by the initial positions. The resulting product is used to increment the coordinates towards the equilibrium configuration. One interesting property of the Hessian matrix is that it is precisely the Dynamical Matrix one uses to compute the phonon spectrum of a vibrating lattice. In our 1DWC system, we can examine the plasma oscillations by simply diagonalizing the Hessian of a relaxed array of particles. Using the interaction potential of Eq.  (\ref{eq:exactint}), the Hessian of a finite size lattice has the form\cite{AM}
\begin{equation}
\mathbf{D(R - R')} = \delta _{\mathbf{R,R'}} \sum\limits_{\mathbf{R''}} {\left. {\frac{{\partial ^2 \phi (x)}}{{\partial x^2 }}} \right|} _{x = \mathbf{R - R''}}
  - \left. {\frac{{\partial ^2 \phi (x)}}{{\partial x^2 }}} \right|_{x = \mathbf{R - R'}} 
  \label{eq:D(R)}
\end{equation}

\begin{figure}
\centerline{\includegraphics[height=2.0in]{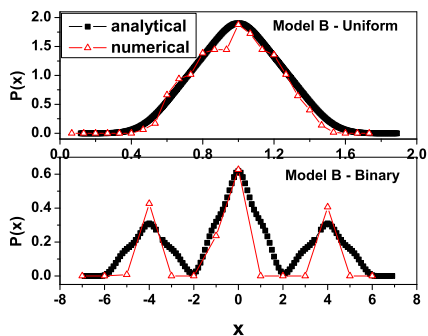}}
\caption{A comparison of the analytical and numerical distribution of nearest neighbor spacings}
\label{fig:unibin}
\end{figure}

\begin{figure}
\centerline{\includegraphics[height=2.0in]{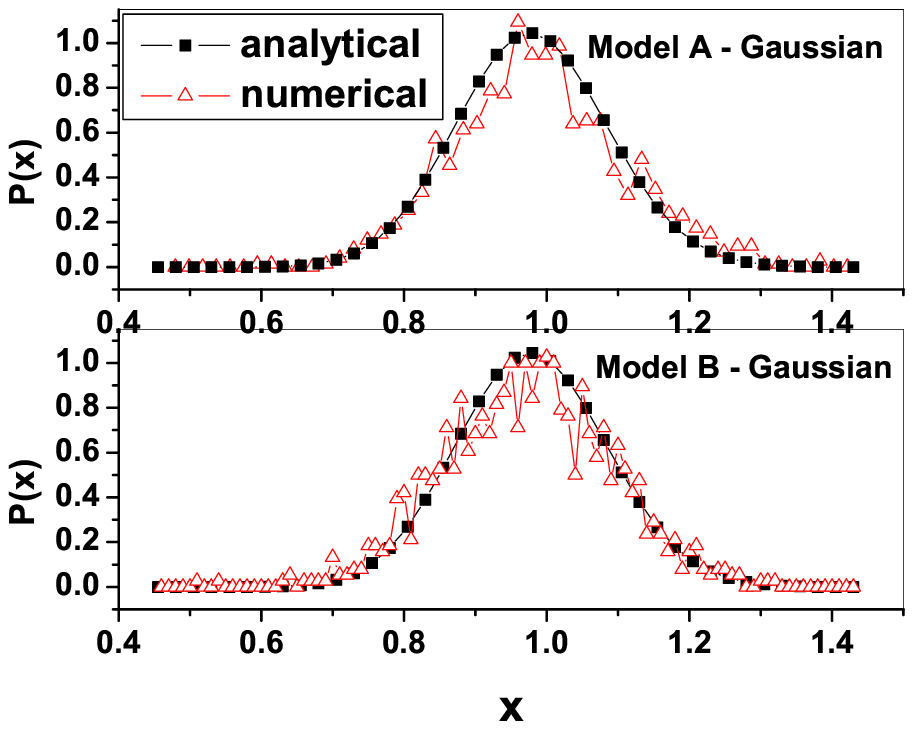}}
\caption{A comparison of the analytical and numerical distribution of nearest beighbor spacings -(continued)}
\label{fig:spacnorm}
\end{figure}

This yields an $N\times N$ matrix that can one use to carry out the standard phonon analysis. We have implemented numerical relaxations on chains of length $N = 1024$, for each of the random charge and random field distributions discussed in the previous section. Numerical uncertainty exists in the amount of residual total force acting on a particular charge. Therefore, the charges were relaxed so that the residual forces were only about $\approx 10^{-6}$ in relative magnitude(scaled to unity), at which point our system is quite close to equilibrium.  We have compared the theoretical predictions of the viscosity approximation given by equations (\ref{eq:unifinal}), (\ref{eq:binfinal}), (\ref{eq:gfinalb}), and (\ref{eq:rpdistfinal}) to the numerical data in Figs. \ref{fig:unibin}, \ref{fig:spacnorm}. We find excellent agreement between the viscosity approximation and numerical calculations in terms in the general behavior of the  resulting distributions. Furthermore, it is clear that the statistical arrangement of the disorder has an explicit dependence on the distribution type and width. The correlation is quite striking in the case of the binary random charge system, where the three characteristic peaks are mirrored in both the numerical and analytical plots. This suggests that an experimental probe of the structural arrangement of disordered systems with long-ranged interactions would reveal a pronounced difference between the chemically disordered binary system and the random potential system, of which represents a form of substrate disorder. It is also worth noting that the effect could also exist in several soft condensed matter models in which Wigner crystal ordering has been discussed in the context of biological systems\cite{boris}.

\section{Crystalline order?}

It is well known that there is a strict absence of long range crystalline order for systems with short-ranged interactions that exist in dimensions $d \leq 2$, resulting from the prevailing influence of thermal and quantum fluctuations. 
This notion has been explored by various authors\cite{2dmelting}.
For Wigner crystals in particular, the unscreened Coulomb interactions factor into these melting criteria through the precise wavevector dependence of the longitudinal and transverse vibrational modes that lead to diverging values of the Lindemann parameter $\left\langle {u^2 } \right\rangle$
in a low enough dimensionality. 

As we have demonstrated in the preceding section, the inclusion of quenched disorder induces randomness in particle spacings. Furthermore, another main consequence of the disorder is that eigenmodes are not plane waves. 
However, as we will show in our investigation of the Bragg peaks, the Fourier space is not necessarily destroyed. 

In our 1DWC system, let us isolate the role of quenched disorder in making the particle array
more amorphous while neglecting the known effects of stochastic zero point disturbances studied by other authors\cite{peeters}. Again, this is valid in systems in which Wigner-like ordering is present but those entropic contributions are suppressed. A more useful route to the exploration of the possibility of long range order involves the structure factor $S_k$, which can be defined in terms of the equilibrium positions of the ith particle, $x_{i}^{equ}$:
\begin{equation}
S_k  = \frac{1}{L}\left\langle {\sum\limits_{i = 1}^L {e^{ikx_i^{equ} } } } \right\rangle ,k = \frac{{2\pi n}}{L},n = 1,2,3 \cdots 
\end{equation}
The crystalline character of the system will manifest itself through the existence of at least one delta function, or Bragg peak, in the spectrum of $S_k$ within some particular scale in Fourier space.

From the exact dispersion relation\cite{1ddisp,schulzwigner:1d_wc} of the clean 1D Wigner crystal, $\omega(k) \sim \left| k \right|\log ^{1/2} (1/k)$, it has been shown that the correlation functions decay much slower than
any power law, $\left\langle {\left[ {u(x) - u(0)} \right]^2 } \right\rangle  \propto \sqrt {\ln (x)}$. The resulting prediction of strong Break peaks in a
scattering experiment follows from the scattering intensity tending like, $I(x)\propto Exp[ - A\left\langle {\left[ {u(x) - u(0)} \right]^2 } \right\rangle ]$, of which
is relatively slow compared to the typical, strong exponential decay observed in systems with short-ranged interactions. According to the literature, systems possessing this sort of decay are usually labeled as having ''quasi-long range order"\cite{bragg}.

The quantity $S_k$ was calculated for numerically relaxed ensembles in both models A and B at different relative disorder strengths. In the case of Model A, the disorder strength strength is the dimensionless combination $Q^2/J$ for values of $J$ defined in Eq. (\ref{eq:rpfourier})), while for model B, it is simply the single charge distribution width $W = 1+ W_0(W_2 - W_1)$, where $W_0 = 0$ is the ordered case. 

\begin{figure}
\centerline{\includegraphics[height=2.0in]{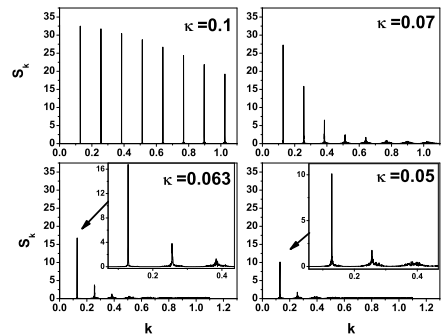}}
\caption{$S_k$ for different relative interaction strengths, model A, L=256.
At high $\kappa$, or low disorder strength, in the delocalization critical regime, the system maintains the composition of the first Bragg peak,
indicating the preservation of some sort of crystalline aspect to the system. }
\label{fig:sfacrf}
\end{figure}

\begin{figure}
\centerline{\includegraphics[height=2.0in]{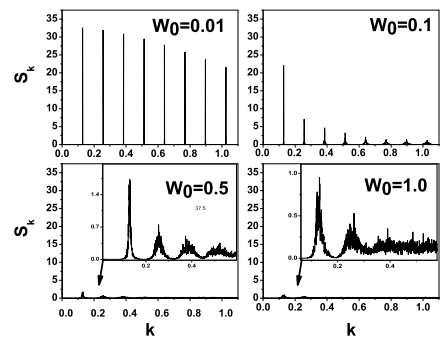}}
\caption{$S_k$ for different disorder strengths, model B, L=256.}
\label{fig:sfacrc}
\end{figure}

The $S_k$ data is shown in Figs. \ref{fig:sfacrf} and \ref{fig:sfacrc}. The plots suggest a difference between both models in terms of their structural properties. It appears that both models retain their crystalline composition
for a regime of disorder strengths. Moreover, in the high disorder regime,
the Bragg peaks are lost. The details of the transition will be examined more carefully in a future work. 
A notable difference between the systems is that in model A the Bragg peaks strongly resemble delta functions while in model B, the scattering intensity contains an exponentially broadened shoulder tending on both sides of the peak at higher disorder strengths, although the presence of peak shoulders in Model B is not evident at low disorder. Hence for weak disorder both models are crystals
based on the strictest definition requiring the existence of just one Bragg peak.

\begin{figure}
\centerline{\includegraphics[height=2.0in]{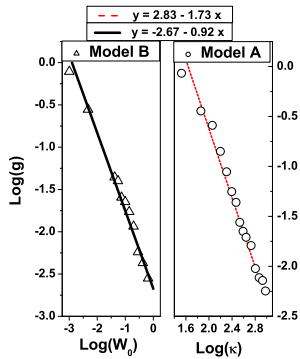}}
\caption{The behavior of the height of the first Bragg peak $g$ for different disorder strengths.}
\label{fig:peakh}
\end{figure}

In order to appropriately make sense of Figs. \ref{fig:sfacrf} and \ref{fig:sfacrc} one should focus on the first Bragg peak and how it behaves as the disorder strength is increased. Evidently, the peak height decreases for both models as function of disorder strength. It is also worth noting that the peaks decrease across Fourier space according to a Debye-Waller damping envelope, tending like,
$I_{DW}  \sim \exp \left( { - \left\langle {(qu(0))^2 } \right\rangle } \right)$.
The precise reduction of the first Bragg peak height $g$ was studied for both models and we determined a power law behavior of the peak height in terms of the disorder strength. We performed a linear fit shown in Fig. \ref{fig:peakh} on a logarithmic scale yielding $ g_A(\kappa ) \sim \kappa ^{ - \alpha } $ and $ g_B(W_0 ) \sim W_0 ^{ - \beta } $, 
for values of $\alpha \approx -1.7272 \pm 0.0696 $ and $\beta \approx -0.9217 \pm 0.0324$. Apparently, this power law behavior breaks down at higher
disorder strengths, as the system becomes more amorphous.

\section{Plasma Oscillations of the Disordered System}

The localization properties of the plasma eigenstates of models A and B have been explored in detail by the authors of this paper \cite{mobility:1d_wc}, where they reported a vibrational delocalization transition at lower eigenfrequencies in Model A, while no transition observed in Model B.  Additionally, they performed a finite size scaling analysis that yielded a scaling behavior of the relative interaction strength $Q^2/J$ and an approximate value of the correlation length exponent. In this section we will focus on other aspects of the plasma eigenstates. In particular, we will explore multifractal properties of the participation moments, the electronic compressibility of the pure and disordered systems, and the frequency dependent behavior of the AC conductivity.

An important physical quantity associated with spatially localized eigenfunctions is the participation ratio $P_q^j$ \cite{wegner:1d_wc} of a given moment $q$, defined as:
\begin{equation}
P_q^j  = \frac{1}{{\sum\nolimits_{k = 1}^L {\left| {u_k^j } \right|^{2q} } }}
\label{eq:pratio}
\end{equation}
where $u_k^j$ is the amplitude of the $jth$ plasma eigenmode at site $k$.

For this particular quantity one must account for the numerical and statistical uncertainties on finite length chains that scale as $1/\sqrt{L}$. One can smooth out the fluctuations of $P_q^j$ by summing over all the eigenstates and ensemble-averaging over several realizations of disorder(represented
by the overline bar). This yields the following length dependent quantity for finite-sized systems:
\begin{equation}
\overline {P_q(L) }  = \frac{1}{L}\overline {\sum\nolimits_{j = 1}^L {P_q^j } } 
\label{eq:avepratio}
\end{equation}

For the 3D Anderson Model the critical scaling properties of $\overline{P_q(L)}$
were derived by Wegner\cite{wegner:1d_wc} using renormalization group arguments in a $2+ \epsilon$ expansion of the 2 particle spectral function of a system of interacting $Q$ matrices. The common view is that all models exhibiting an Anderson transition will possess critical eigenfunctions that can be used to construct the quantity $\overline{P_q(L)}$, having a length dependence that scales like: 
\begin{figure}
\centerline{\includegraphics[height=2.0in]{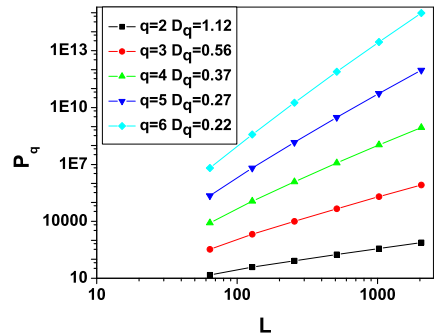}}
\caption{Averaged participation moments showing multifractal scaling, Model A}
\label{fig:ipm}
\end{figure}
\begin{equation}
\overline {P_q  (L)} \propto L^{D_q (q - 1)} 
\end{equation}
or in other words, $\overline {P_q  (L)}$ exhibits multifractal scaling, in that one obtains a different exponent $D_q$ for each moment $q$. Other authors such as Lima \emph{et. al.} have related this scaling behavior to the wave-packet dynamics of the power-law bond disordered one-dimensional Anderson model with hopping amplitudes decreasing as $H_{nm} \propto \left| n - m \right|^{-\alpha}$. They were able to extract a time-dependent scaling analysis of the participation moments by performing a finite size scaling analysis of the electronic return probability\cite{lima:1d_wc}. Furthermore, they report an asymptotic value of $D_{\infty} = 0.5$. We have performed a similar calculation shown in Fig. \ref{fig:ipm}. Indeed, the critical eigenfunctions of Model A exhibit multifractal behavior. However the asymptotic value $D_{\infty} \approx 0.2$, differs from the value 0.5 obtained in the analogous electronic tightbinding system.

It is important to note that the behavior of the plasma oscillations have important consequences for physical quantities such as the electron compressibility $\kappa$. We remark that our definition of the 
$\kappa$ only includes the electron gas itself and does not consider the compensating background. The main reason for using this definition is that the 
background contribution does not enter into the experimentally determined values of $\kappa$ when using the double well field penetration technique, of which has been widely applied to measure the compressibility of interacting electron systems in semiconductor heterostructure systems\cite{Eisenstein}. The thermodynamic compressibility can be related to the volume fluctuations of a system around its average value, given by the relation
\begin{equation}
\kappa \propto \frac{{\left\langle {\left( {\Delta V} \right)^2 } \right\rangle }}{{\left\langle V \right\rangle }}
\end{equation}
One can associate the normal modes $u(x)$ with these volume fluctuations in a finite system by the relation,
\begin{equation}
\kappa  \propto \frac{1}{L}\left\langle {\sum\limits_{n = 1}^L {\frac{1}{{\omega _n^2 }}\left( {\left. {\frac{{\partial u}}{{\partial x}}} \right|_{x_n } } \right)^2 } } \right\rangle 
\label{eq:fincomp}
\end{equation}
where $\omega _n$ is the nth eigenfrequency and $x_n$ labels the nth oscillator.

For the ordered---or clean---Wigner crystal, the precise behavior of the compressibility can be extracted by exploiting the spatial periodicity of the longitudinal plasma eigenfrequencies $\omega_L$ at long wavelengths\cite{martin}.
\begin{equation}
\kappa \propto \mathop {\lim }\limits_{q \to 0} \frac{{q^2 }}{{\omega _L^2 (q)}}
\label{eq:sumrule}
\end{equation}
Apparently, for the classical ordered system one can substitute the long wavelength form of the dispersion relations $\omega(k) \sim \left| k \right|\log ^{1/2} (1/k)$,   into (\ref{eq:sumrule}) to yield a vanishing compressibility, $\kappa = 0$\cite{1ddisp}. 

Similarly, the quantum calculations having the dispersion relation\cite{gold:1d_wc}
\begin{equation}
\omega _{RPA} (k) \approx \frac{{2e_0 }}{{\sqrt \pi  }}\sqrt {v_f } \left| k \right|\log ^{1/2} \left( {\frac{1}{{kd}}} \right)
\label{eq:rpadisp}
\end{equation}
also yield $\kappa = 0$, where $v_f$ is the Fermi velocity and $d$ is a finite system width.

 We also note that our definition of the compressibility does not always yield a vanishing result in the thermodynamic limit. For a harmonic oscillator system with random couplings and nearest neighbor interactions we observe a finite compressibility. This appears to be consistent with a dispersion $\omega(k)$ that is linear in $k$, in the ordered system, of which, $\kappa=0$ via Eq.  (\ref{eq:sumrule}). This leads us to believe that the compressibility of the charges is dependent on the range of the interactions, not the nature of the disorder.

We have also applied Eq.  (\ref{eq:fincomp}) to Models A and B at various system sizes and have observed a vanishing compressibility $\kappa = 0$, in the limit of a large system, for both cases. This is a significant result, given that the properties of disordered elastic systems has recently been subject of intense research\cite{ledoussal:1d_wc}. We are confident that the methods used in this article can easily be extended to two dimensional systems, where a major unresolved a problem is the compressibility behavior observed in 2D SI MOSFETS. Various research teams have experimentally  observed a vanishing compressibility on the insulating side of the well known 2D Metal-Insulator transition\cite{jiang:1d_wc},\cite{allison}.  

There has been much speculation as to whether the insulating phase is an incompressible zero field Wigner crystal. Certain authors have suggested the use of the gaussian variational replica method(GSM) as a viable tool for understanding disordered Wigner crystals. However, it is known that these elastic models assume a continuum limit that washes out the full microscopic structural arrangement of the electrons; for example one could not discuss the distribution of nearest neighbor spacings in a random charge system like Model B using the GSM, given that the spacings between nearest neighbor particles is quantity observed only by a discrete description as we have considered in this paper. Moreover, even if one wishes to apply a continuum description, one needs to determine the elastic constants to effectively utilize the GSM method for computing the relevant physical quantities, thus rendering previous replica based studies inconclusive.  On the other hand, we have been able to calculate the compressibility by circumventing these field theoretical limitations on elastic systems(constrained by the Larkin scale) through the use of simple classical phonon methods. We hope to discuss the 2D calculations in a future article, given the fact that we have also developed some precise mathematical tools for handling the 2D long range Coulomb interactions, of which are analogous to the exact summation in Eq.  (\ref{eq:exactint}).
	
Another important quantity that derives from the plasma oscillations is the AC conductivity. The plasmon propagator $G(\omega)$ is itself the direct response to an external electric field interacting with the electrons. $G(\omega)$ can be determined from the Hessian or dynamical matrix, defined by Eq.  (\ref{eq:D(R)}), as constructed in the following resolvent:
\begin{equation}
G(\omega ) = \sum\limits_R^L {\sum\limits_{R'}^L {\frac{1}{{\omega ^2 \mathbb{I} - D(R,R')}}} } 
\label{eq:gfun}
\end{equation}
The AC conductivity is then given by the quantity\cite{AM}:
\begin{equation}
\sigma (\omega ) = iA\omega G(\omega )
\label{eq:accond}
\end{equation}

\begin{figure}
\centerline{\includegraphics[height=2.0in]{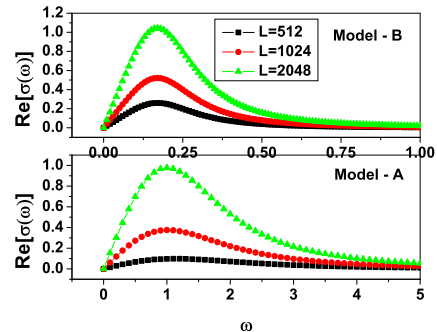}}
\caption{$\mathop{\rm Re}[\sigma (\omega)]$ vs. $\omega$ for a fixed relative disorder strength in both models $\kappa = 0.1$, $W_0 = 0.5$ at different system sizes.}
\label{fig:acfsize}
\end{figure}

\begin{figure}
\centerline{\includegraphics[height=2.0in]{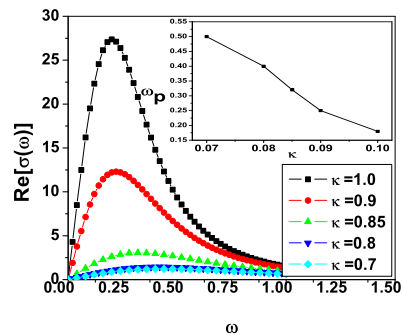}}
\caption{$\mathop{\rm Re}[\sigma(\omega)]$ vs. $\omega$ at different relative disorder strength values, $\kappa$(L=512).
The inset shows the shift of the pinning frequency $\omega_p$ as the disorder is decreased. }
\label{fig:acintscal}
\end{figure}

The effects of disorder on the zero temperature AC transport properties was discussed by Giamarchi and Schultz\cite{giamschulz:1d_wc} for 1D electron systems and Chitra et al. for 2D Wigner Crystals\cite{zerofield:1d_wc}. With reference to this earlier work we have calculated the AC conductivity via Eq.  (\ref{eq:accond}). The finite size effects are shown in Fig.\ref{fig:acfsize} for both Models A and B. The key quantities of interest are the peak height $P$ and its associated frequency $\omega_p$.   For the pure crystalline system $\omega_p$ corresponds to the Drude driving frequency. The disorder acts as a broadening mechanism, by which $\omega_p$ is shifted to a new frequency known as the pinning frequency and also the peak height $P$ is shifted as well. We have qualitatively studied how $\omega_p$ scales with the relative interaction strength of the Coloumb potential over the random potential previously defined as $\kappa$ in the critical Model A. With reference to our earlier work, the observed interaction scaling in Model A can be utilized to relate 
the critical eigenfrequency at which delocalization takes place $\omega_c$ to the pinning frequency $\omega_p$. We have plotted the behavior of $\mathop{\rm Re}[\sigma(\omega)]$ at various interaction strengths shown in Fig. \ref{fig:acintscal}. There is a noticeable increase in peak height with increasing interaction strength and a subtle decrease in $\omega_p$ as displayed in the window of Fig. \ref{fig:acintscal}. Consequently, this implies that for critical frequencies $\omega_c$ closer to the upper band edge there is a smaller pinning frequency. This can be further interpreted as connecting stronger plasma eigenmode localization with higher pinning frequencies. We remark that more careful scaling studies, both analytical and numerical must performed for a deeper understanding of the connection between critical depinning transitions and vibrational delocalization. 

\begin{figure}
\centerline{\includegraphics[height=2.0in]{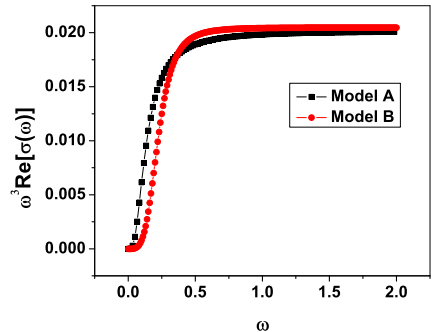}}
\caption{$\mathop{\rm \omega^3 Re}[\sigma (\omega)]$ vs. $\omega$ for a fixed relative disorder strength in both models $\kappa = 0.1$, $W_0 = 0.5$ at L=512. The zero slope regime at higher frequencies
clearly suggests a $1/\omega^3$ dependence.}
\label{fig:highac}
\end{figure}

Another important property of the AC conductivity is the high frequency behavior of $\mathop{\rm Re}[\sigma (\omega)]$. We examined the limit $\omega >> \omega_p$ and observed an $\omega$ dependence going like $\mathop{\rm Re}[\sigma (\omega)]\sim 1/\omega^3$ for both Models A and B. This is shown in Fig. \ref{fig:highac}. Certain authors have determined a universal $\omega^4$ dependence at low frequencies for 1D disordered Wigner Crystals while others have established a $\omega^{2K-4}$ for high frequencies from renormalization group studies of interacting 1D electron systems, where $K$ is an effective scaling coupling constant for repulsive interactions\cite{fogler:1d_wc}. These earlier studies only consider short ranged interactions while our system takes into account the true long-ranged nature of the Coulomb interactions. The latter form would apply to our classical systems for the value $K \ll 1$, yielding a $1/\omega^4$ dependence. For the purposes of comparison with similar 1D disordered electron models, our dependence on $\omega$ differs, for our results are precisely the same as the 2D system studied by Giamarchi et al. using the GSM technique, where they also observe $\mathop{\rm Re}[\sigma (\omega)]\sim 1/\omega^3$ behavior for high frequencies. 
	  
\section{Conclusion and Experimental Prospects}

In this paper we have considered a wide range of phenomena in 1D disordered Wigner solids that follow from the behavior of the plasma oscillations.  We have developed a new mathematical formalism for studying the spatial arrangement of the particles at equilibrium as a consequence of a certain type
of randomness in the Hamiltonian. We have applied probability convolutions to quantify the distribution of nearest neighbor particle spacings by assuming the particle coordinates will deviate from crystalline configuration by distances that are proportional to the total force on each particle when the system is
ordered. If one adiabatically ``turns on'' the randomness, one can imagine the charges propagating in a viscous medium, and if the random contribution to the forces on the charges is small a linear relaxation of the particles might be expected to hold as a good approximation.  We have tested this viscosity approximation against numerically relaxed ensembles, and observe what appears to be remarkably good agreement. A stronger statistical analysis will be performed in future papers that will further develop this probability convolution formalism on more rigorous grounds.

We have also explored the basic question of whether a 1DWC in the presence of quenched disorder retains its crystalline composition. We have argued that both a random potential system and a random charge system generate an internal stiffness that resists the system from completely melting in the thermodynamic limit, at least when the effect of randomness is not too great. 

We briefly remark that a melting transition has been observed at a critical value of disorder strength. However, this will be formally reported in a future article.  

In the process of studying Models A and B numerically, we have revised existing computational paradigms for incorporating long ranged Coulomb forces in systems with periodic boundary conditions and disorder. A numerical relaxation scheme has been outlined along with a procedure for studying the collective modes of disordered Wigner crystals or other ``glasslike'' structures in general.  

We conclude by emphasizing the necessity for experimental probes of the density response functions in order that one may investigate the possibility of a vibrational delocalization transition and our predictions of a zero compressibility phase. Glasson \emph{et. al.} have already observed Wigner Crystal ordering in a quasi-1D system\cite{wire:1d_wc}. We consider the possibility of well-controlled manipulation of the substrate randomness as a viable avenue by which the scaling behavior of models A and B can be investigated experimentally. Furthermore, based on the observed connection between the delocalization transition and the pinning frequency $\omega_p$, AC conductivity measurements are also quite essential for elucidating our understanding of the basic physics involved.

\begin{acknowledgements}
We thank M.M. Fogler for comments on the manuscript.
In addition, we acknowledge useful discussions with S. Chakravarty and S.E. Brown.
\end{acknowledgements}

\appendix

\section{The distribution of random charge products}
\label{sec:appA}

In general, the charge product density function is given as a convolution of two different densities\cite{stat}:
\begin{equation}
P(t) = \int {\int {P_{q,i}(Q_i } } )P_{q,j}(Q_j )\delta (Q_i Q_j  - t)dQ_i dQ_j 
\end{equation}
If the two charges share a common density function, then:
\begin{equation}
P(t) = \int {P_q(Q)P_q(\frac{t}{Q})\frac{{dQ}}{\left|Q\right|}} 
\end{equation}
The range of values for $t$ must be carefully determined for various regimes, depending on the type and width of distribution $P_q(Q_i)$. We start by examining the case of a uniform distribution of random single charges in Model B, in which
\begin{equation}
P_q(Q)= \left\{ \begin{array}{ll}
 \frac{1}{{W_2  - W_1 }} & W_1  < Q < W_2  \\ 
 0 & {\rm otherwise} \\ 
 \end{array} \right.
 \label{eq:unidist}
\end{equation}
Evidently, nucleons and electrons can only have integer values of the charge $e$, so this particular distribution is unphysical and unrealistic.  We consider this distribution only as a version of a randomly-distributed coupling constant and for the purposes of determining which features of the system are robust with respect to distribution type.

After determining the range of values for $t$, we can solve for the joint distribution for a product of charges:
\begin{equation}
P(t) = \left\{ \begin{array}{ll}
 \frac{{\ln \left| {t/W_1 ^2 } \right|}}{{(W_2  - W_1 )}} & t < W_2 W_1  \\ 
 \frac{{\ln \left| {W_2 ^2 /t} \right|}}{{(W_2  - W_1 )}} & t > W_2 W_1  \\ 
 \end{array} \right.
\label{eq:uniforce}
\end{equation}
As a more germane alternative, we also consider a binary distribution for Model B, defined as simply a sum of two delta functions with equal statistical weights. This particular type of disorder can be interpreted as a form of chemical disorder or a type of binary alloy. We consider the binding of chemical constituents to the Wigner crystal such that there are effectively two values of charges random distributed throughout the system: 
\begin{equation}
P_q(Q) = \frac{1}{2}\delta (Q - Q_1 ) + \frac{1}{2}\delta (Q - Q_2 )
\label{eq:bindist}
\end{equation}
The product distribution simply consists of three peaks at the values $Q_1^{2}$, $Q_1 \times Q_2$, and $Q_2^{2}$.
\begin{equation}
P(t) = \frac{{\rm{1}}}{{\rm{4}}}\delta (Q_1^2  - t) + \frac{{\rm{1}}}{{\rm{2}}}\delta (Q_1 Q_2  - t) + \frac{{\rm{1}}}{{\rm{4}}}\delta (Q_2^2  - t)
\label{eq:binforce}
\end{equation}

Finally, we consider the gaussian product distribution:
\begin{equation}
P(t) = \int\limits_{ - \infty }^\infty  {\int\limits_\infty ^\infty  {e^{ - (Q_i-\mu) ^2 /2\sigma _i } } } e^{ - (Q_j-\mu) ^2 /2\sigma _j } 
( Q_i  Q_j  - t)d Q_i dQ_j
\label{eq:gforce} 
\end{equation}
For $\mu =0$ this integral can be evaluated exactly to yield
\begin{equation}
P(t) = (\frac{1}{{\pi \sigma ^2 }})K_0 \left( {\frac{{\left| t \right|}}{{\sigma ^2 }}} \right)
\end{equation}
where $K_0$ is a modified Bessel function of the second kind.\cite{stegun}
For a non-zero mean the product distribution can be determined numerically.

\section{The Ewald summation of a quasi-1D periodic system}
\label{sec:appB}

We carry out the Ewald summation technique for a quasi-1D system of charges in a uniform neutralizing background. The electrostatic potential of interest has the following form:
\begin{equation}
\frac{1}{{\left| {r - nL} \right|}} - \int {\frac{{dx/L}}{{\sqrt {(r - x)^2  + b^2 } }}} 
\end{equation}
The first term can be expanded out as an integral:
\[
\frac{1}{{\left| {r - nL} \right|}} = \int_0^\infty  {e^{ - t(r - nL)^{2}} } t^{ - 1/2} dt
\]
This integral can be broken up into two parts by introducing a suitable cutoff $\frac{{\pi \beta }}{{L^2 }}$:
\[
 = \int_0^{\frac{{\pi \beta }}{{L^2 }}} {e^{ - t(r - nL)^{2}} } t^{ - 1/2} dt + \int_{\frac{{\pi \beta }}{{L^2 }}}^\infty  
{e^{ - t(r - nL)^{2}} } t^{ - 1/2} dt
\]
The 2nd integral can be evaluated through a variable substitution and can be left as is:
\[
t \to \frac{{\pi x}}{{L^2 }} \Rightarrow \sum\limits_{n =  - \infty }^\infty  {\int_\beta ^\infty  {\frac{{\sqrt{\pi}}}{{L}}
e^{ - \pi x(\frac{r}{L} - n)^2 } } x^{ - 1/2} dx} 
\]
Next, we apply the Poisson sum formula to the 1st integral.
\[
 = \sum\limits_{m =  - \infty }^\infty  {\int_0^{\frac{{\pi \beta }}{{L^2 }}} {\int\limits_{ - \infty }^\infty  {e^{2\pi imn} }
 e^{ - t(r - nL)} } t^{ - 1/2} dndt} 
\]
The integral on $n$ can carried out by a standard completion of the squares trick yielding,
\[
 = \sum\limits_{m =  - \infty }^\infty  {\frac{{\sqrt \pi  }}{L}\int\limits_0^{{\raise0.7ex\hbox{${\pi \beta }$} \!\mathord{\left/
 {\vphantom {{\pi \beta } {L^2 }}}\right.\kern-\nulldelimiterspace}
\!\lower0.7ex\hbox{${L^2 }$}}} {e^{(2\pi irm/L) - (\pi ^2 m^2 /L^2 t)} } } \frac{{dt}}{t}
\]
It is important that we properly subtract the divergent contribution of the $m=0$ term. The resulting summand contains a lower incomplete gamma function:
\[
 = \sum\limits_{m \neq 0 }^\infty  {e^{(2\pi irm/L)} } \Gamma (0,\frac{{m^2 \pi }}{\beta })
\]
Lastly we subtract the divergent contribution at $m=0$ arising from the neutralizing background,
\[
\int {\frac{{dx/L}}{{\sqrt {(r - x)^2  + b^2 } }}}  = \int\limits_0^\infty  {\int {\frac{{dx}}{L}} } e^{ - [(r - x)^2  + b^2 ]t} t^{ - 1/2} dt
\]
If we include the other $m=0$ term from the previous summand the contributions to the sum become,
\[
\int\limits_0^{\pi \beta /L^2 } {(1 - e^{b^2 t} } )\frac{{dt}}{t} + \int\limits_{\pi \beta /L^2 }^\infty  {e^{b^2 t} } \frac{{dt}}{t} 
\approx \ln \beta  + C
\] 
where C is some constant. Our final expression for the long-ranged electrostatic potential becomes,
\begin{equation}
\begin{aligned}
V(r) = \sum\limits_{n =  - \infty }^\infty  {\int\limits_\beta ^\infty  {e^{ \pi x(\frac{r}{L} - n)^2 } x^{1/2}dx }}\\
+ \sum\limits_{m \ne 0} {e^{2\pi imr/L}   \Gamma (0,\frac{{m^2 \pi }}{\beta })} -  \ln (\beta )
\label{eq:ewaldf}
\end{aligned}
\end{equation}
It is important to note that the expression converges to a value that is independent of the cutoff $\beta$.

\end{document}